\documentclass[aps,prb,twocolumn,superscriptaddress,a4paper]{revtex4-1}
\usepackage[dvipdfmx]{graphicx}

\usepackage{color}
\usepackage{amsmath}
\usepackage{amssymb}
\usepackage[pdftex,colorlinks=true,linkcolor=blue,citecolor=blue,filecolor=blue,baseurl=empty]{hyperref}

\begin{document}

\title{Exceptional points in the one-dimensional Hubbard model}

\author{Roman Rausch}
\email[]{r.rausch@tu-braunschweig.de}
\address{Technische Universit\"at Braunschweig, Institut f\"ur Mathematische Physik, Mendelssohnstra{\ss}e 3,
38106 Braunschweig, Germany}
\address{Department of Physics, Kyoto University, Kyoto 606-8502, Japan}

\author{Robert Peters}
\address{Department of Physics, Kyoto University, Kyoto 606-8502, Japan}

\author{Tsuneya Yoshida}
\address{Department of Physics, University of Tsukuba, Ibaraki 305-8571, Japan}

\begin{abstract}
Non-Hermitian phenomena offer a novel approach to analyze and interpret spectra in the presence of interactions. Using the density-matrix renormalization group (DMRG), we demonstrate the existence of exceptional points for the one-particle Green’s function of the 1D alternating Hubbard chain with chiral symmetry, with a corresponding Fermi arc at zero frequency in the spectrum. They result from the non-Hermiticity of the effective Hamiltonian describing the Green’s function and only appear at finite temperature. They are robust and can be topologically characterized by the zeroth Chern number. This effect illustrates a case where temperature has a strong effect in 1D beyond the simple broadening of spectral features. Finally, we demonstrate that exceptional points appear even in the two-particle Green’s function (charge structure factor) where an effective Hamiltonian is difficult to establish, but move away from zero frequency due to a distinct symmetry constraint.
\end{abstract}

%\noindent{\it Keywords\/}: non-Hermitian, Hubbard model, density-matrix renormalization group

\maketitle

%%%%%%%%%%
\section{\label{sec:Introduction}Introduction}
%%%%%%%%%%

The Hermiticity of a Hamiltonian that results in real, measurable eigenenergies, is one of the fundamental assumptions of quantum mechanics. Still, over the years it became clear that studying non-Hermitian Hamiltonians is also worthwhile, as they
are relevant for certain physical situations. This is particularly obvious for open and nonequilibrium systems~\cite{Fukui_1998,Ashida_2020}, where the energy is not conserved, including optical cavities~\cite{Feng_2017,Ozawa_2019,Oezdemir_2019} or cold atoms with particle losses~\cite{Yamamoto_2019,Yoshida_2020c}. However, the concept of non-Hermiticity enters even into closed equilibrium systems via an effective-Hamiltonian description that may result from interactions~\cite{Kozii_2017,Yoshida_2018,Kimura_2019,Matsushita_2019,Michishita_2020,Michishita_2020b,Yoshida_2020}
or disorder~\cite{Zyuzin_2018,Shen_2018,Papaj_2019,Matsushita_2020}.

In particular, photoemission and inverse photoemission experiments of correlated systems are related to the single-particle Green's function $G(\omega,\ensuremath{\mathbf{k}})=\left[\omega-H_0\left(\ensuremath{\mathbf{k}}\right)-\Sigma\left(\omega,\ensuremath{\mathbf{k}}\right)\right]^{-1}$, where $H_0\left(\ensuremath{\mathbf{k}}\right)$ is the noninteracting Hamiltonian and $\Sigma\left(\omega,\ensuremath{\mathbf{k}}\right)$ is the self-energy, a function of frequency $\omega$ and momentum $\ensuremath{\mathbf{k}}$. The Green's function is thus being governed by an effective Hamiltonian~\cite{Kozii_2017} $H_{\mathrm{eff}}\left(\omega,\ensuremath{\mathbf{k}}\right)=H_0\left(\ensuremath{\mathbf{k}}\right)+\Sigma\left(\omega,\ensuremath{\mathbf{k}}\right)$, which is in general a non-Hermitian matrix due to the imaginary part of the self-energy that describes the damping of quasiparticles.

A principal property of non-Hermitian matrices is that they can become non-diagonalizable at the so-called ``exceptional points''. 
The bulk spectrum of the effective Hamiltonian at such a point shows a novel topological band touching,
which can be characterized by vorticity~\cite{Shen_2018b} (or, equivalently, a winding number). Furthermore, exceptional points induce Fermi arcs, along which the bandgap becomes purely imaginary.
The topological aspect of this band touching can be studied by taking symmetries into account and it is possible to find higher-dimensional exceptional rings and surfaces~\cite{Shen_2018,Okugawa_2019,Budich_2019,Kawabata_2019b,Yoshida_2020}. Apart from exceptional points, non-Hermiticity induces a new arena of other topological phenomena~\cite{SanJose_2016,Gong_2018,Kawabata_2019b,Bergholtz_2019,Yoshida_2019c,Liu_2020}.
This may, for example, result in an unusual bulk-boundary correspondence~\cite{Yao_2018,Kunst_2018,Zhang_2019,Lee_2019,Okuma_2020,Yokomizo_2019,Borgnia_2020,Helbig_2020,Hofmann_2020,Xiao_2020,Yoshida_2020b,Yoshida_2020c}.

Unlike noninteracting topological insulators, however, we stress that a key requirement for the novel non-Hermitian phenomena are lifetime effects, which may stem from interactions, disorder or the coupling to a bath. In this way, they form a bridge between topology and strongly correlated quantum systems~\cite{Kozii_2017,Yoshida_2018}. It also means that, when analyzing the non-Hermitian aspects of interacting systems, one faces the inevitable hurdle of having to solve an intractable many-body problem. Therefore, despite the enormous progress in this field, previous works were based on severe approximations such as a momentum-independent self-energy $\Sigma\left(\omega,\ensuremath{\mathbf{k}}\right) \approx \Sigma\left(\omega\right)$ or even a constant self-energy $\Sigma\left(\omega,\ensuremath{\mathbf{k}}\right) \approx i\gamma$. Furthermore, previous works were limited to an analysis of the single-particle Green's function.  

In this paper, we demonstrate the existence of exceptional points in a strongly correlated 1D system and their effect on the one-particle properties. By using the numerically exact density matrix renormalization group (DMRG), the self-energy includes full momentum dependence and no drastic approximations beyond numerical cutoffs are employed.

Our results show that a pair of exceptional points emerges at the endpoints of a 1D Fermi arc in the one-particle Green's function at finite temperature. They appear due to chiral (sublattice) symmetry and can thus be characterized by the zeroth Chern number~\cite{Yoshida_2019} (a zero-dimensional topological invariant).

Moreover, DMRG allows us to extend the scope beyond one-particle excitations, so that we are able to show how non-Hermiticty in a strongly correlated system affects two-particle observables, where an effective-Hamiltonian description is not easily obtainable. In particular, we demonstrate the emergence of exceptional points in the two-particle Green's function and Fermi arcs in the dynamical structure factor. In contrast to the single-particle Green's function, the exceptional points emerge away from the Fermi energy even in the presence of chiral symmetry. These distinct behaviors are due to the fact that the many-body symmetry imposes a different symmetry constraint on each Green's function.

The presentation of our results is structured as follows: Section \ref{sec:1D} pedagogically discusses the general conditions for the emergence of exceptional points in 1D systems, which is followed by section \ref{sec:Model}, where we establish a minimal model based on these criteria. We discuss one-particle properties at zero temperature in section \ref{sec:T0}, and at finite temperatures in section \ref{sec:T}. In section \ref{sec:Topology}, we present the topological characterization of exceptional points emerging at finite temperatures and explicitly demonstrate the robustness of the Fermi arc in section \ref{sec:Robustness}. Finally, we analyze the two-particle Green's function in section \ref{sec:G2P} before concluding our findings. 

%%%%%%%%%%%%%%%%%%
%\section{\label{sec:Results}Results}
%%%%%%%%%%%%%%%%%%

%%%%%%%%%%%%%%%%%%%%%
\section{\label{sec:1D}Prerequisites}
%%%%%%%%%%%%%%%%%%%%%

As mentioned in the introduction, the retarded single-particle Green's function, given by $G\left(\omega,\ensuremath{\mathbf{k}}\right) = \big[\omega-H_0\left(\ensuremath{\mathbf{k}}\right)-\Sigma\left(\omega,\ensuremath{\mathbf{k}}\right)\big]^{-1}$ (where $H_0\left(\ensuremath{\mathbf{k}}\right)$ is the noninteracting Hamiltonian and $\Sigma\left(\omega,\ensuremath{\mathbf{k}}\right)$ is the self-energy), is governed by the effective Hamiltonian
\begin{equation}
H\textsubscript{eff}\left(\omega,\ensuremath{\mathbf{k}}\right) = H_0\left(\ensuremath{\mathbf{k}}\right) + \Sigma\left(\omega,\ensuremath{\mathbf{k}}\right),
\end{equation}
which is in general non-Hermitian if the system is interacting and there is a finite quasiparticle lifetime $\mathrm{Im}\Sigma\left(\omega,\ensuremath{\mathbf{k}}\right) \neq 0$~\cite{Kozii_2017}.

The minimal model to observe exceptional points in the one-particle Green's function has two sublattices (equivalently, two bands), so that $H\textsubscript{eff}$  is a $2\times2$ matrix. It can be written in the basis of Pauli matrices $\boldsymbol{\tau}=\left(\tau_1,\tau_2,\tau_3\right)$ and the identity matrix $\tau_0$ with complex coefficients $c_i = b_i+id_i$ ($b_i\in\mathbb{R}$, $d_i\in\mathbb{R}$):
\begin{equation}
H\textsubscript{eff} = \left(b_0+id_0\right) \tau_0 + \left(\ensuremath{\mathbf{b}}+i\ensuremath{\mathbf{d}}\right) \cdot \boldsymbol{\tau}.
\label{eq:Heff_general}
\end{equation}
The eigenvalues are given by
\begin{equation}
E_{\pm} = b_0+id_0 \pm \sqrt{b^2-d^2+2i\ensuremath{\mathbf{b}}\cdot\ensuremath{\mathbf{d}}},
\label{eq:Epm}
\end{equation}
and the eigenvectors by
\begin{equation}
v_{\pm} = \frac{1}{\sqrt{N} }
\biggl( \begin{array}{c}
c_3 \pm \sqrt{b^2-d^2+2i\ensuremath{\mathbf{b}}\cdot\ensuremath{\mathbf{d}}} \\
c_1+ic_2
\end{array} \biggr),
\end{equation}
where $\sqrt{N}$ is a normalization prefactor. Whenever the two conditions
\begin{eqnarray}
f\left(\omega\textsubscript{EP},\ensuremath{\mathbf{k}}\textsubscript{EP}\right) &=& b^2-d^2 = 0 \label{eq:cond_bdsq},\\
g\left(\omega\textsubscript{EP},\ensuremath{\mathbf{k}}\textsubscript{EP}\right) &=& \ensuremath{\mathbf{b}}\cdot\ensuremath{\mathbf{d}} = 0, \label{eq:cond_bd}
\end{eqnarray}
are satisfied, the square root vanishes, the eigenvalues become degenerate, and there is only one independent eigenvector. At this  ``exceptional point'' ($\omega\textsubscript{EP}$,$\ensuremath{\mathbf{k}}\textsubscript{EP}$), the effective Hamiltonian becomes non-diagonalizable. Since the two above conditions are linearly independent, one needs at least a two-dimensional parameter space to satisfy them without fine-tuning. In particular, when focusing on the experimentally relevant Fermi energy $\omega=0$, this implies that a two-dimensional momentum space $\left(k_x,k_y\right)$ is needed~\cite{Kozii_2017, Yoshida_2018, Yoshida_2020}.

\begin{figure}[t]
\centering
\begin{center}
\includegraphics[width=\columnwidth]{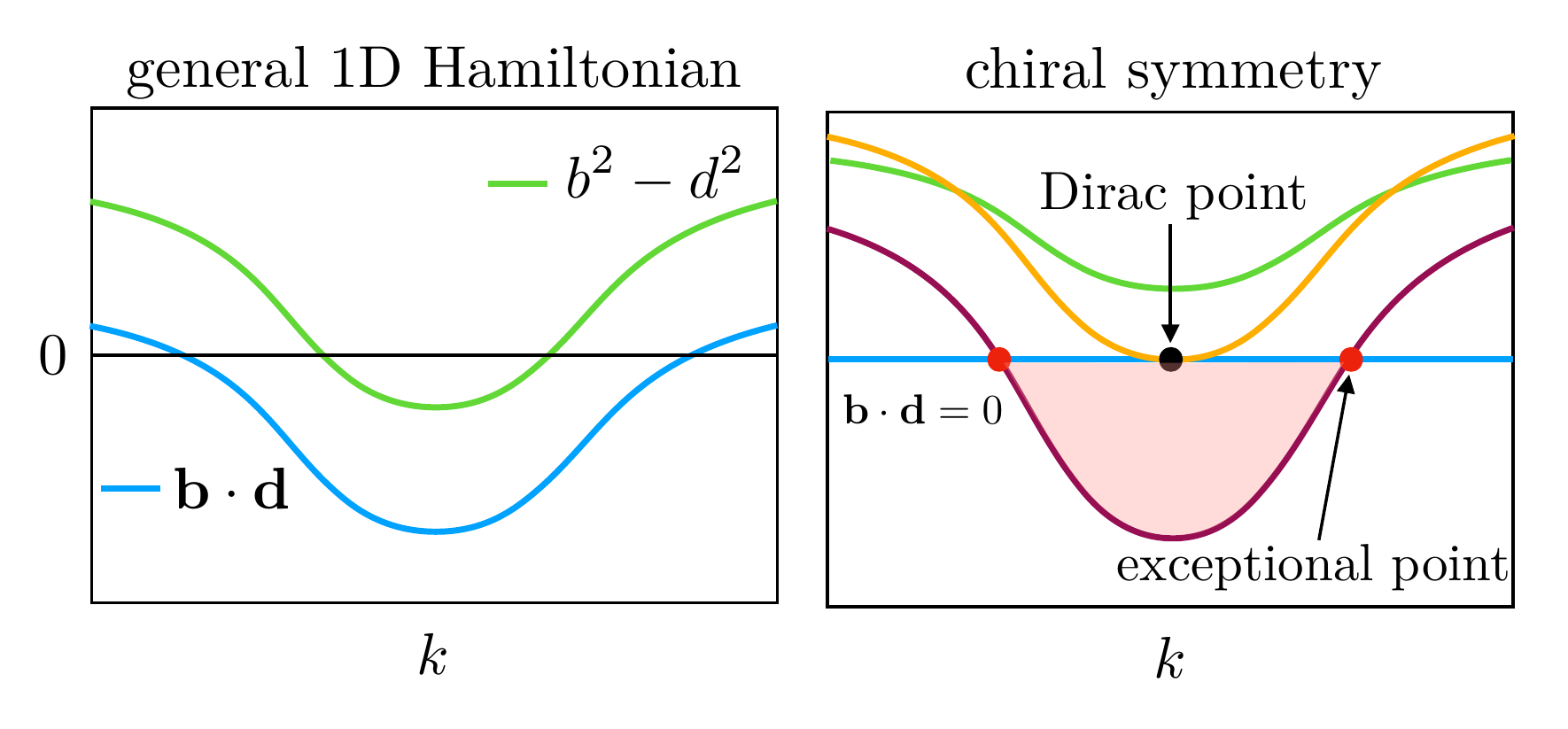}
\caption{\label{fig:chiralSymm}
Left: Sketch of the behavior of $f\left(\omega=0,k\right) = b^2-d^2$ and $g\left(\omega=0,k\right) = \ensuremath{\mathbf{b}}\cdot\ensuremath{\mathbf{d}}$ for a generic 1D Hamiltonian. The functions may cross zero, but not at the same point without fine-tuning the system parameters.
Right: The same for a 1D Hamiltonian with chiral symmetry, where $\ensuremath{\mathbf{b}}\cdot\ensuremath{\mathbf{d}}$ is fixed to zero, while $f\left(\omega=0,k\right) = b^2-d^2$ may touch zero at the Dirac point (yellow curve) or cross zero at the exceptional points (purple curve).
}
\end{center}
\end{figure}

Alternatively, it is possible to restrict one condition by symmetry~\cite{Yoshida_2020}. This can be seen in the following way: Suppose that a 1D effective Hamiltonian at the Fermi energy $H\textsubscript{eff}\left(k\right)=H\textsubscript{eff}\left(\omega=0,k\right)$ satisfies the chiral (or sublattice) symmetry given by
\begin{equation}
\tau_3 H^{\dagger}\left(k\right)\tau_3 = -H\left(k\right).
\label{eq:chiralSymm}
\end{equation}
It implies that $b_0=0$, $\ensuremath{\mathbf{b}}=\left(b_1,b_2,0\right)$ and $\ensuremath{\mathbf{d}}=\left(0,0,d_3\right)$, so that condition~(\ref{eq:cond_bd}) is always satisfied. The eigenvalues now reduce to
\begin{equation}
E_{\pm} = id_0 \pm \sqrt{b_1^2+b_2^2-d_3^2}.
\label{eq:EpmChiral}
\end{equation}
Thus, the zeros of the periodic function $f\left(k\right)=b_1^2+b_2^2-d_3^2$ correspond to exceptional points. In the trivial case, it has no zeros at all. In the non-trivial case, it may touch zero in one point or cross zero in an even amount of points.
This is illustrated in figure~\ref{fig:chiralSymm}. 
We note that $d_0$ and $d_3$ are only nonzero if an imaginary part of the self-energy is present. Hence, $f\left(k\right)$ is a positive semi-definite function in the noninteracting case, being a sum of two squares. Turning on an interaction which is equal in both sublattices $\Sigma_{00}=\Sigma_{11}$ adds an imaginary part to the effective Hamiltonian, but only contributes to $d_0$, so that $f\left(k\right)$ remains positive semi-definite.
On the other hand, an interaction that is sublattice-dependent, $\Sigma_{00}\neq\Sigma_{11}$, also contributes to $d_3$ and may lead to zeros in $f\left(k\right)=b_1^2+b_2^2-d_3^2$.

Furthermore, looking at Eq.~(\ref{eq:EpmChiral}) we note that in the region where $f\left(k\right)=b_1^2+b_2^2-d_3^2<0$, the energy eigenvalues become purely imaginary, indicating that the two bands coalesce into one broadened line segment at the Fermi energy that stretches between the exceptional points, a one-dimensional analogue of a Fermi arc. As such, the non-Hermitian nature of the effective Hamiltonian induced by interaction has a drastic effect on the band structure that can be observed in the experiment. In the following, we construct a minimal model to observe this effect and prove the existence of the 1D Fermi arc.

%%%%%%%%
\section{\label{sec:Model}Model}
%%%%%%%%

The necessary ingredients to observe the exceptional points as described in the previous section are a system with chiral symmetry and a difference in self-energies of the sublattices. A not strictly necessary, but very helpful condition is the presence of a Dirac point at the Fermi energy, where the bands cross in the noninteracting case. At this point, $H_0\left(k\right)$ vanishes, but $f\left(k\right)$ only touches zero (see figure~\ref{fig:chiralSymm}). The addition of an arbitrary small self-energy contribution is then expected to split this Dirac point into two exceptional points.

The minimal model that satisfies these conditions is given by the alternating Hubbard Model
\begin{equation}
\begin{split}
H &= -\sum_{ij\sigma} t_{ij} \left(c^{\dagger}_{i\sigma} c_{j\sigma} +H.c.\right) \\
&\quad+\sum_i U_i \left(n_{i\uparrow}-\frac{1}{2}\right)\left(n_{i\downarrow}-\frac{1}{2}\right),
\label{eq:H}
\end{split}
\end{equation}
with
\begin{equation}
  t_{ij} = \left.
   \begin{cases}
    t_0 & \textrm{if } i, j \textrm{ are nearest neighbors,}\\
    0 & \textrm{ else,}\\
  \end{cases}
  \right.
\end{equation}
and
\begin{equation}
  U_i = \left.
  \begin{cases}
    U_A & \textrm{for } i \textrm{ even,}\\
    U_B & \textrm{for } i \textrm{ odd,}\\
  \end{cases}
  \right.
\end{equation}
where $c^{\dagger}_{i\sigma}$ creates an electron with the spin projection $\sigma = \uparrow, \downarrow$ at the site given by the coordinate $R_i$ and $n_{i\sigma}=c^{\dagger}_{i\sigma}c_{i\sigma}$ is the corresponding density. The ground state of the model is found at half filling $N = \sum_{i} \left<n_{i}\right> = \sum_{i\sigma} \left<n_{i\sigma}\right> = L$, where $L$ is the length of the 1D chain. The hopping amplitude $t_0\equiv1$ sets the energy scale, and with $\hbar\equiv 1$ also the time scale.
This model can be interpreted as a one-dimensional chain with two different, alternating atoms, labeled $A$ and $B$.

In the simplest case, we have $U_A\neq0$ and $U_B=0$. Such a model was previously investigated for its various properties that are different from the homogeneous chain: transfer of the magnetic momentum to the free sites~\cite{Paiva_1996}, the appearance of a giant magnetoresistance effect~\cite{Li_2018, Baibich_1988}, a Mott insulator transition that may occur at fillings other than half filling~\cite{Paiva_1998}, and the formation of a modulated and potentially incommensurate charge-density wave~\cite{Paiva_2002, Malvezzi_2006, Zhang_2015}.
However, we note that in contrast to previous studies, our model includes alternating on-site energies instead of a homogeneous chemical potential. In this way, a charge-density-wave is suppressed, and the ground state is found at half-filling for each site ($\left<n_i\right>=1$), rather than at half-filling averaged over a unit cell.

By writing the coordinates $R_i=mL_c+R_{\mu}$ with $m\in\mathbb{Z}$, $L_c=2$ being the length of the unit cell, and $R_{\mu}=0,1$ for $\mu=A,B$ within the unit cell; and Fourier-transforming $c_{\mu}\left(k\right) = 1/\sqrt{L/L_c} \sum_m \exp\left(-ikmL_c\right) c_{m\mu}$ between the cells, we can obtain the effective Hamiltonian at the Fermi energy in units of $t_0$ as
\begin{equation}
H\textsubscript{eff}\left(k\right) =
\biggl( \begin{array}{cc}
i\mathrm{Im}\Sigma_A\left(0,k\right) & -1-e^{-ik}\\
-1-e^{ik} & i\mathrm{Im}\Sigma_B\left(0,k\right)
\end{array}
\biggl),
\end{equation}
from which we can read off the coefficients of Eq.~(\ref{eq:Heff_general}) as $b_1\left(k\right)=-1-\cos\left(k\right)$ , $b_2\left(k\right)=\sin{k}$, $d_0\left(k\right)=\Gamma_+\left(k\right)$, $d_3\left(k\right)=\Gamma_-\left(k\right)$ with $\Gamma_{\pm}\left(k\right) = 1/2 \big[\mathrm{Im}\Sigma_A\left(\omega=0,k\right)\pm\mathrm{Im}\Sigma_B\left(\omega=0,k\right)\big]$. The noninteracting system has a Dirac cone at $k=\pi$ (see figure~\ref{fig:2_T0}).

%%%%%%%%%%
\section{\label{sec:T0}Zero temperature}
%%%%%%%%%%

We calculate the one-particle retarded Green's function at zero temperature (see appendix~\ref{app:Methods}) defined as:
\begin{equation}
\begin{split}
G_{m\mu,n\nu}^{1p}\left(t\right) = 
&-i\theta\left(t\right)\sum_{\sigma} \left<0\big|e^{iHt} c_{m\mu\sigma} e^{-iHt} c^{\dagger}_{n\nu\sigma}\big|0\right>\\
&-i\theta\left(t\right)\sum_{\sigma} \left<0\big|e^{-iHt} c^{\dagger}_{m\mu\sigma} e^{iHt} c_{n\nu\sigma}\big|0\right>,
\label{eq:G_T=0}
\end{split}
\end{equation}
where $\theta\left(t\right)$ is the step function (taking $0$, $1/2$ and $1$ for $t<0$, $t=0$ and $t>0$, respectively) and $\left|0\right>$ is the ground state. It is then Fourier-transformed between cells to yield:
\begin{equation}
G^{1p}_{\mu\nu}\left(\omega,k\right) = \int_0^{\infty} dt~ e^{i\omega t} \sum_{nm} G_{m\mu,n\nu}\left(t\right) e^{-ik\left(m-n\right)L_c}.
\label{eq:G_FT}
\end{equation}

Around the Dirac point, exceptional points should already appear for weak coupling, but the smaller $U_A$, the more difficult they are to resolve. Throughout the paper we therefore set $U_A=4$, which is is in the intermediate-coupling regime, being equal to the noninteracting bandwidth $W=4$.

\begin{figure*}[!htb]
\centering
\begin{center}
\includegraphics[width=0.7\textwidth]{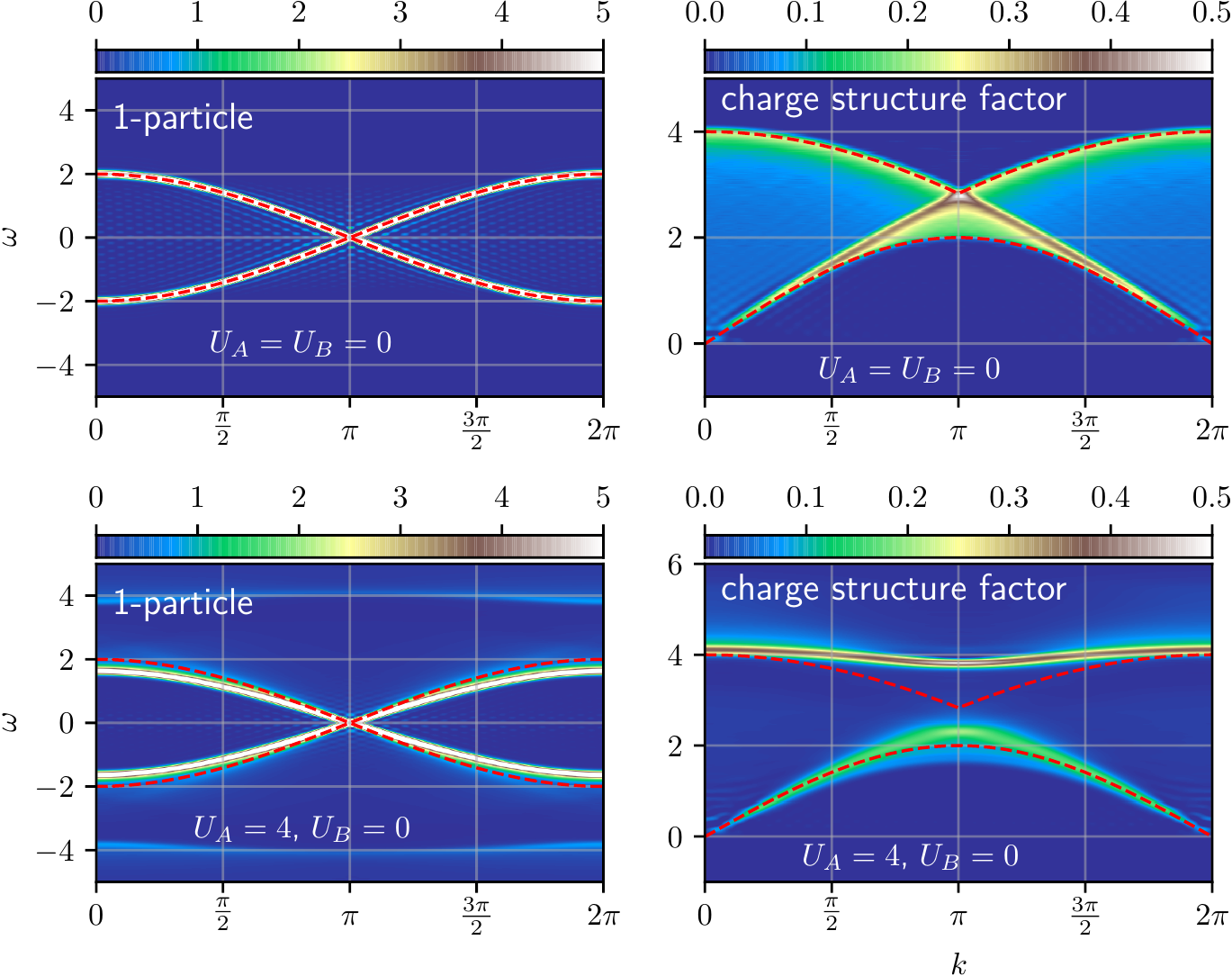}
\caption{\label{fig:2_T0}
Left: Trace of the spectral function $-1/\pi \mathrm{Im} \mathrm{Tr} G^{1p}\left(\omega,k\right)$ from the one-particle Green's function (\ref{eq:G_T=0}) and Eq.~(\ref{eq:G_FT}) at zero temperature (see appendix~\ref{app:Methods} for technical details). The upper figure is for $U_A=U_B=0$, the lower figure for $U_A=4$, $U_B=0$. The red dotted line shows the analytical free dispersion.
Right: The same for the two-particle charge-charge Green's function, first part of (\ref{eq:Gcc2}) for $T=0$, where the red dotted line now shows the boundaries of the two-electron continuum.
}
\end{center}
\end{figure*}

The left part of figure \ref{fig:2_T0} shows the result.  We notice that this interaction introduces Hubbard bands separated by about $U_A$ with a small spectral weight, and otherwise only slightly renormalizes the bands crossing at the Dirac point, leaving the cone in place.
Thus, we can conclude that  the imaginary part of the self-energy vanishes at the Fermi energy for zero temperature. While such a behavior is guaranteed by the Fermi liquid theory in 3D, it does not hold in general for 1D systems. In our case, we can understand it as a consequence of setting $U_B=0$, which leads to a mixed behavior: Hubbard bands due to a finite $U_A$, but no change around the Fermi energy due to $U_B=0$.

The consequence is that the observation of 1D non-Hermitian effects at the Fermi energy requires an additional condition which creates a finite lifetime $\mathrm{Im}\Sigma(\omega=0,k)\neq0$. We find that setting $U_B>0$ alone does not help, since apart from creating a strong self-energy, it immediately causes a gap with vanishing spectral weight around $\omega=0$. Instead, the effect we are looking for can be found by going to finite temperatures.

%%%%%%%%%%%%%%%%
\section{\label{sec:T}Finite temperature}
%%%%%%%%%%%%%%%%

When working with finite temperature, we have to switch from a description by a wavefunction to a density matrix. Using the purification formalism~\cite{Feiguin_2005, Karrasch_2013, Barthel_2016, Nocera_2016}, it can be flattened to a vector $\big|\beta\big>=e^{-\beta H/2}\big|\beta=0\big>$, where $\beta=1/T$ is the inverse temperature. We now have to calculate
\begin{equation}
\begin{split}
G^{1p}_{m\mu,n\nu}\left(t\right) 
&= -i\theta\left(t\right)Z\left(\beta\right)^{-1}\\
&\quad\times\bigg[ \left<\beta\big|e^{-iHt} c^{\dagger}_{m\mu\sigma} e^{iHt} c_{n\nu\sigma}\big|\beta\right>\\
&\quad+\left<\beta\big|e^{iHt} c_{m\mu\sigma} e^{-iHt} c^{\dagger}_{n\nu\sigma}\big|\beta\right>
\bigg],
\end{split}
\label{eq:G_finiteT}
\end{equation}
where $Z\left(\beta\right)=\left<\beta\big|\beta\right>$ is the partition function. See appendix~\ref{app:Methods} for more technical details.

\begin{figure*}[!htb]
\centering
\begin{center}
\includegraphics[width=0.8\textwidth]{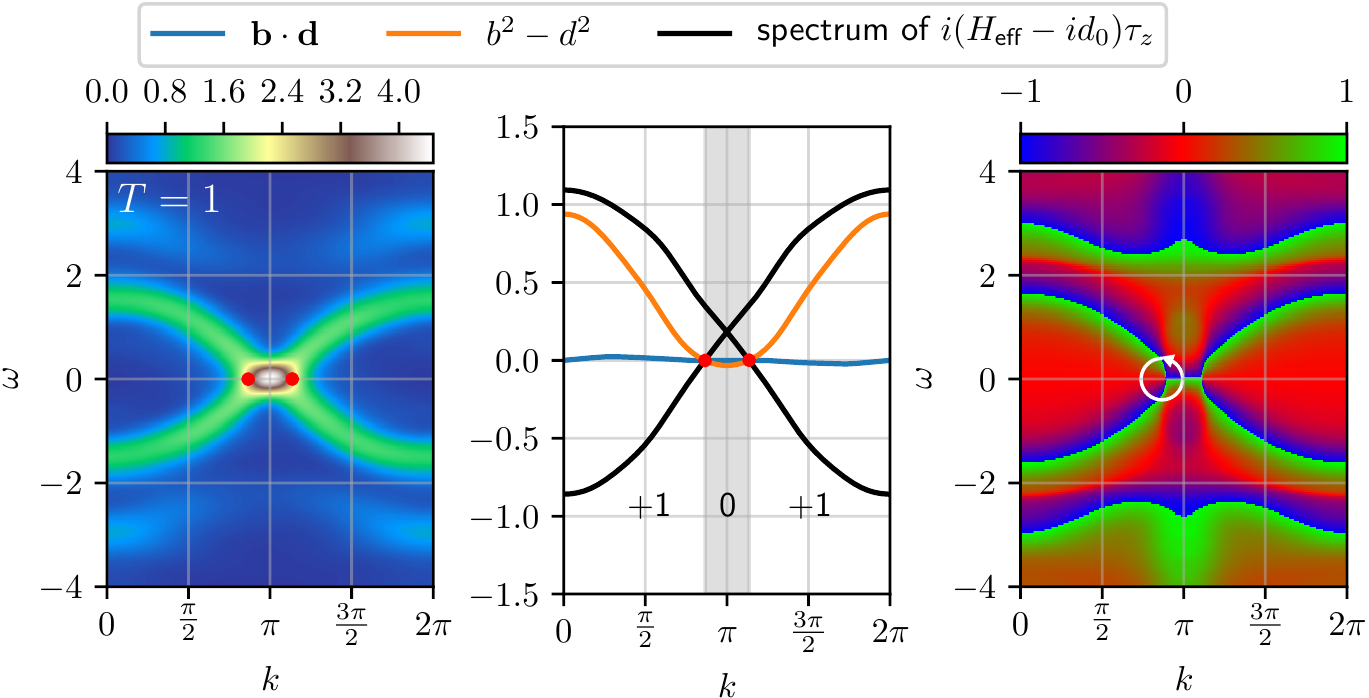}
\caption{\label{fig:1pT}
Left: Trace of the one-particle spectral function $-1/\pi \mathrm{Im} \mathrm{Tr} G^{1p}\left(\omega,k\right)$ from Eq.~(\ref{eq:G_finiteT}) and Eq.~(\ref{eq:G_FT}) calculated for $T=1/\beta=1$ (see appendix~\ref{app:Methods} for technical details). 
Center:  $\ensuremath{\mathbf{b}}\cdot\ensuremath{\mathbf{d}}$ and $b^2-d^2$ calculated from $\left(H\textsubscript{eff}\right)_{\mu\nu}\left(k\right) = -\big[\left(G^{1p}\right)^{-1}\big]_{\mu\nu}\left(\omega=0,k\right)$; as well as the eigenvalues of $H_+\left(k\right)=i\left[H\textsubscript{eff}\left(k\right)-id_0\left(k\right)\right]\tau_3$. The red dots mark the exceptional points, while the area shaded in grey marks the change in number of negative eigenvalues of $H_+$ from 0 to 1.
Right: The corresponding phase function $\phi\left(\omega,k\right)$ Eq.~(\ref{eq:arg}). The white circle illustrates finite vorticity (\ref{eq:vort}) around the first exceptional point.
}
\end{center}
\end{figure*}

The result is shown in the left part of figure~\ref{fig:1pT} for $T=1$. 
We now indeed observe that the Dirac cone splits into two exceptional points with a Fermi arc of large spectral weight in between. 
In the middle panel, we see that $\ensuremath{\mathbf{b}}\cdot\ensuremath{\mathbf{d}}$ vanishes (with only a small numerical error) at the Fermi energy for all momenta as required by chiral symmetry.
Thus, the intersections of $b^2-d^2$ with zero correspond to the exceptional points, marked by red dots. The same dots are shown overlaid on the spectral function. 
The resulting self-energy includes full momentum dependence, as briefly discussed in appendix~\ref{app:Sigma}. The length of the Fermi arc as a function of temperature is shown in figure~\ref{fig:arcLength}. It grows with increasing $T$ and eventually saturates, reaching a total width of about $0.4\pi$ (for the given $U_A=4$, $U_B=0$). The Fermi arc in this model is equivalent to a ``flat band'' located exactly at the Fermi energy and thus has strong effect on observable properties.

%%%%%%%%%%%%%%%%%%%
\section{\label{sec:Topology}Topological characterization}
%%%%%%%%%%%%%%%%%%%

\subsection{Zeroth Chern number with chiral symmetry}

As has been shown in previous works~\cite{Kawabata_2019c,Yoshida_2019, Yoshida_2020}, the exceptional points at $\omega=0$ in presence of chiral symmetry can be characterized by the zeroth Chern number, which is the number of negative eigenenergies of the Hermitian matrix $H_{+}\left(k\right)=i\left[H\textsubscript{eff}\left(k\right)-id_0\left(k\right)\right]\tau_3$.
The spectrum of $H_+\left(k\right)$, as calculated from the DMRG data, is shown as the center plot of figure~\ref{fig:1pT}. We see that the number of negative eigenvalues indeed changes from 1 to 0 at the same points that are obtained from the zeros of $b^2-d^2$, further proving that these anomalies are indeed exceptional points. Because any perturbation of the parameters, either in the Hamiltonian or the temperature, results in a smooth change of the band structure of $H_{+}\left(k\right)$, we can conclude that the exceptional points shown here are robust. This is explicitly proven in the next section.

\subsection{Vorticity in $\omega$-$k$ space}

Furthermore, we may ask the question of what happens if chiral symmetry is broken: Do the exceptional points disappear immediately? We believe that they will survive, but cease to be fixed at $\omega=0$. This can be seen by computing the vorticity in $\omega$-$k$ space, which is well-defined even in the absence of symmetries~\cite{Shen_2018b,Michishita_2020b}. This quantity is related to the complex eigenenergies of the effective Hamiltonian, but because the one-particle Green's function is essentially its inverse, we can simply rewrite the vorticity in terms of the Green's function $G^{1p}$.
Namely, we can expand $G^{1p}$ itself in the basis of Pauli matrices
\begin{equation}
G^{1p}\left(\omega,k\right) = \left(b_0+id_0\right) \tau_0 + \left(\ensuremath{\mathbf{b}}+i\ensuremath{\mathbf{d}}\right) \cdot \boldsymbol{\tau},
\label{eq:G1p_bd}
\end{equation}
and look at the following phase function~\cite{Michishita_2020b}:
\begin{equation}
\phi\left(\omega,k\right) = 1/\pi \arg\left(b^2-d^2+2i\ensuremath{\mathbf{b}}\cdot\ensuremath{\mathbf{d}}\right).
\label{eq:arg}
\end{equation}
The vorticity is then given by 
\begin{equation}
v=\oint \nabla_{\mathbf{r}}\phi\left(\omega,k\right) d\mathbf{r},
\label{eq:vort}
\end{equation}
with $\mathbf{r}=\left(\omega,k\right)$, $\nabla_{\mathbf{r}} := \left(\partial_w, \partial_k\right)$ and the integral is taken around a closed path in the $\left(\omega,k\right)$-plane. 

The phase function $\phi\left(\omega,k\right)$ is shown on the right side of figure~\ref{fig:1pT}. There are discontinuities, which in the topologically trivial case form closed surfaces as $k$ winds around the Brillouin zone. Thus, a closed path around any point encounters an even number of phase jumps and the vorticity vanishes. This is not the case at the exceptional points, where three phase jumps are encountered (shown by the white circle) and a nonzero phase is picked up, which can be seen without calculating $v$ explicitly. We note that introducing a symmetry-breaking perturbation  does not change the value of the vorticity. 
Thus, we can conclude that breaking the chiral symmetry just shifts the exceptional points away from $\omega=0$ line to the two-parameter ($\omega$,$k$)-space, at least for small perturbations. However, investigating this effect in more detail is beyond the scope of the present paper.

%%%%%%%%%%%%%%%%%%%%%
\section{\label{sec:Robustness}Robustness of the Fermi arc}
%%%%%%%%%%%%%%%%%%%%%

The robustness of the exceptional points and the Fermi arc can be verified by directly perturbing the Hamiltonian. A one-particle perturbation of this kind is given by a dimerized hopping
\begin{equation}
  t_{ij} = \left.
  \begin{cases}
    t_0\left(1-\delta/2\right)=:t_- & \textrm{for } i \textrm{ even; } j=i+1 \textrm{ odd,}\\
    t_0\left(1+\delta/2\right)=:t_+ & \textrm{for } i \textrm{ odd; } j=i+1 \textrm{ even,}\\
  \end{cases}
  \right.
\label{eq:t_dimer}
\end{equation}
and the effective Hamiltonian in units of $t_0$ becomes:
\begin{equation}
H^{\delta}\textsubscript{eff}\left(k\right) =
\biggl( \begin{array}{cc}
i\mathrm{Im}\Sigma_A\left(0,k\right) & -t_--t_+e^{-ik}\\
-t_--t_+e^{ik} & i\mathrm{Im}\Sigma_B\left(0,k\right)
\end{array} \biggr) 
.
\end{equation}

\begin{figure*}[!htb]
\centering
\begin{center}
\includegraphics[width=0.6\textwidth]{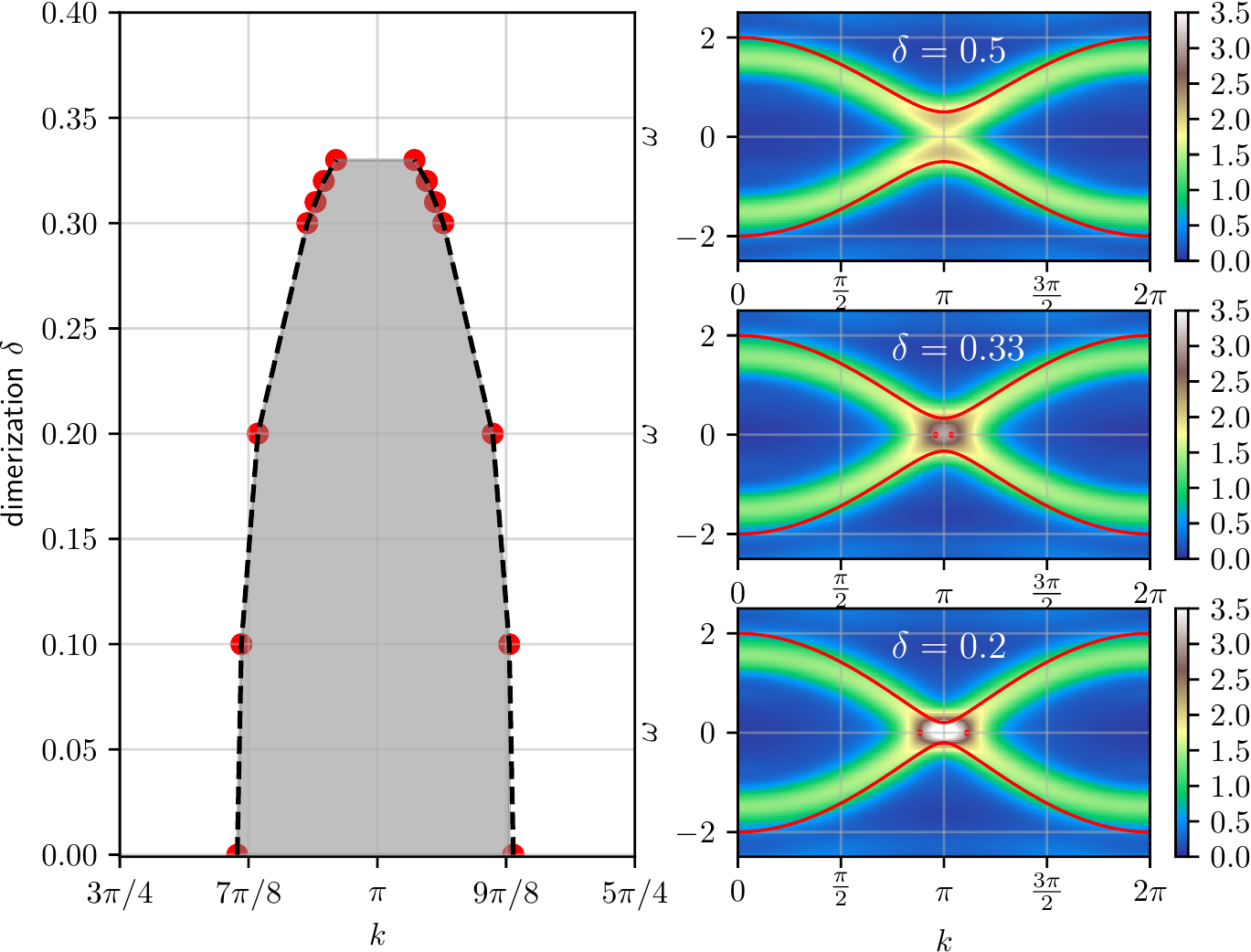}
\caption{\label{fig:dimerization}
Left: Position of the exceptional points in the one-particle spectrum for various values of the dimerization $\delta$ (see Eq.~(\ref{eq:t_dimer}) and~(\ref{eq:H})), for $T=1$, $U_A=4$, $U_B=0$.
Right: The corresponding Fermi arc and spectral weight for various values of $\delta$ as shown. The red lines indicate the free dispersion where a Peierls transition takes place for any $\delta>0$.
}
\end{center}
\end{figure*}

In the noninteracting case, $\delta$ causes a Peierls transition, with a gap appearing for any finite $\delta>0$ in one spatial dimension. This is different from the interacting case displayed in figure~\ref{fig:dimerization}. Small dimerizations almost do not affect the gapless Fermi arc at all. For $\delta\gtrsim0.2$, it starts to shrink and eventually disappears as the exceptional points pair-annihilate at $\delta_c\sim0.335$ for $U_A=4$, $U_B=0$, and a gap eventually opens in the spectrum.

\begin{figure*}[!htb]
\centering
\begin{center}
\includegraphics[width=0.45\textwidth]{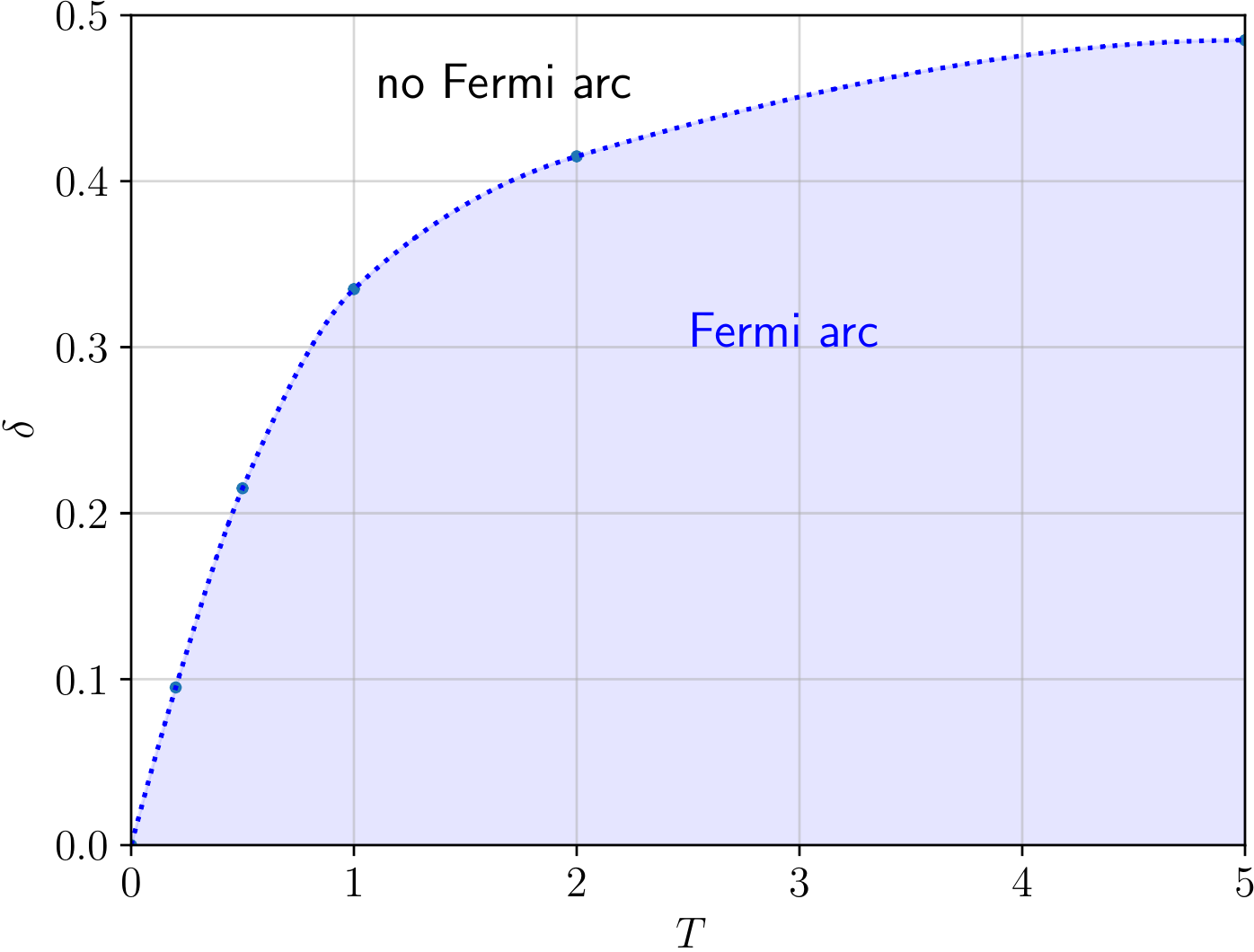}
\caption{\label{fig:phasediag}
Critical dimerization $\delta_c$ at which the Fermi arc disappears as a function of temperature $T$. Parameters as in figure~\ref{fig:dimerization}.
}
\end{center}
\end{figure*}

Figure \ref{fig:phasediag} shows the corresponding ``phase diagram'', i.e. the position of the critical value $\delta_c$ where the Fermi arc vanishes as a function of the temperature for the same fixed values of $U_A=4$, $U_B=0$. 
One can conclude that if one is interested in observing the Fermi arc, the disruptive effect of a strong dimerization can be compensated by a higher temperature (which increases the imaginary part of the self-energy). However, for the given interaction strength, a dimerization that exceeds $\delta\approx0.5$ cannot be overcome. Thus, even though the exceptional points can be annihilated by strong dimerization, they are stable against fairly high values of $\delta$.

We have also checked that a small $U_B>0$ has much the same effect as dimerization, namely the Fermi arc remains robust up to a certain critical value $U_{B,c}$. Altogether, this indicates the presence of a ``Fermi arc phase'' in the 4-parameter space spanned by $U_A$, $U_B$, $T$ and $\delta$. Computing its precise boundaries within this space is beyond the scope of the current paper and is left for future investigations.

%%%%%%%%%%%%%%%%%%%%%%%%%%%%%
\section{\label{sec:G2P}Two-particle spectral function}
%%%%%%%%%%%%%%%%%%%%%%%%%%%%%

Finally, we investigate how the non-Hermitian effects present themselves in the two-particle spectral function. Due to the unitary spin and charge SU(2) symmetries of our Hamiltonian, the only independent local two-particle excitations are given by the charge density operator $n_i=\sum_{\sigma}n_{i\sigma}$ and the spinflip operator $S^+_{i}=c^{\dagger}_{i\uparrow}c_{i\downarrow}$ (see appendix~\ref{app:Unitary}). However, we find that the results for both are very similar (as finite temperature destroys any tendency of spin or charge order), so that for reasons of brevity we concentrate only on the charge excitations. By convention we use the pseudospin operator
\begin{equation}
Q^z_i=1/2\left(n_{i}-1\right),
\label{eq:Qz}
\end{equation}
and calculate the charge-charge Green's function defined via the commutator
\begin{eqnarray}
G^{cc}_{m\mu,n\nu}\left(t\right) &=
-i\theta\left(t\right)Z\left(\beta\right)^{-1}\left<\beta\big|\big[Q^z_{m\mu\sigma}\left(t\right),Q^z_{n\nu\sigma}\big]\big|\beta\right>.
\label{eq:Gcc1}
\end{eqnarray}
Using $\left<XY\left(t\right)\right>=\left<X\left(-t\right)Y\right> = \left<Y^{\dagger}\left(-t\right)X^{\dagger}\right>$ for general operators $X$ and $Y$, we can write it in the following, numerically more convenient, form:
\begin{equation}
\begin{split}
G^{cc}_{m\mu,n\nu}\left(t\right) &=
-i\theta\left(t\right)Z\left(\beta\right)^{-1} \\
&\quad\times\bigg[ \left<\beta\big|e^{iHt} Q^z_{m\mu\sigma} e^{-iHt} Q^z_{n\nu\sigma}\big|\beta\right>\\
&\quad-\left<\beta\big|e^{-iHt} Q^z_{m\mu\sigma} e^{iHt} Q^z_{n\nu\sigma}\big|\beta\right>
\bigg].
\label{eq:Gcc2}
\end{split}
\end{equation}
Note that just like the one-particle Green's function consists of two parts that correspond to photoemission and inverse photoemission in the experiment, the two-particle Green's function also has two parts, albeit with a relative minus sign. The first is the dynamical charge structure factor (CSF), the second one could be called the ``inverse charge structure factor'' (ICSF).

\begin{figure*}[!htb]
\centering
\begin{center}
\includegraphics[width=0.8\textwidth]{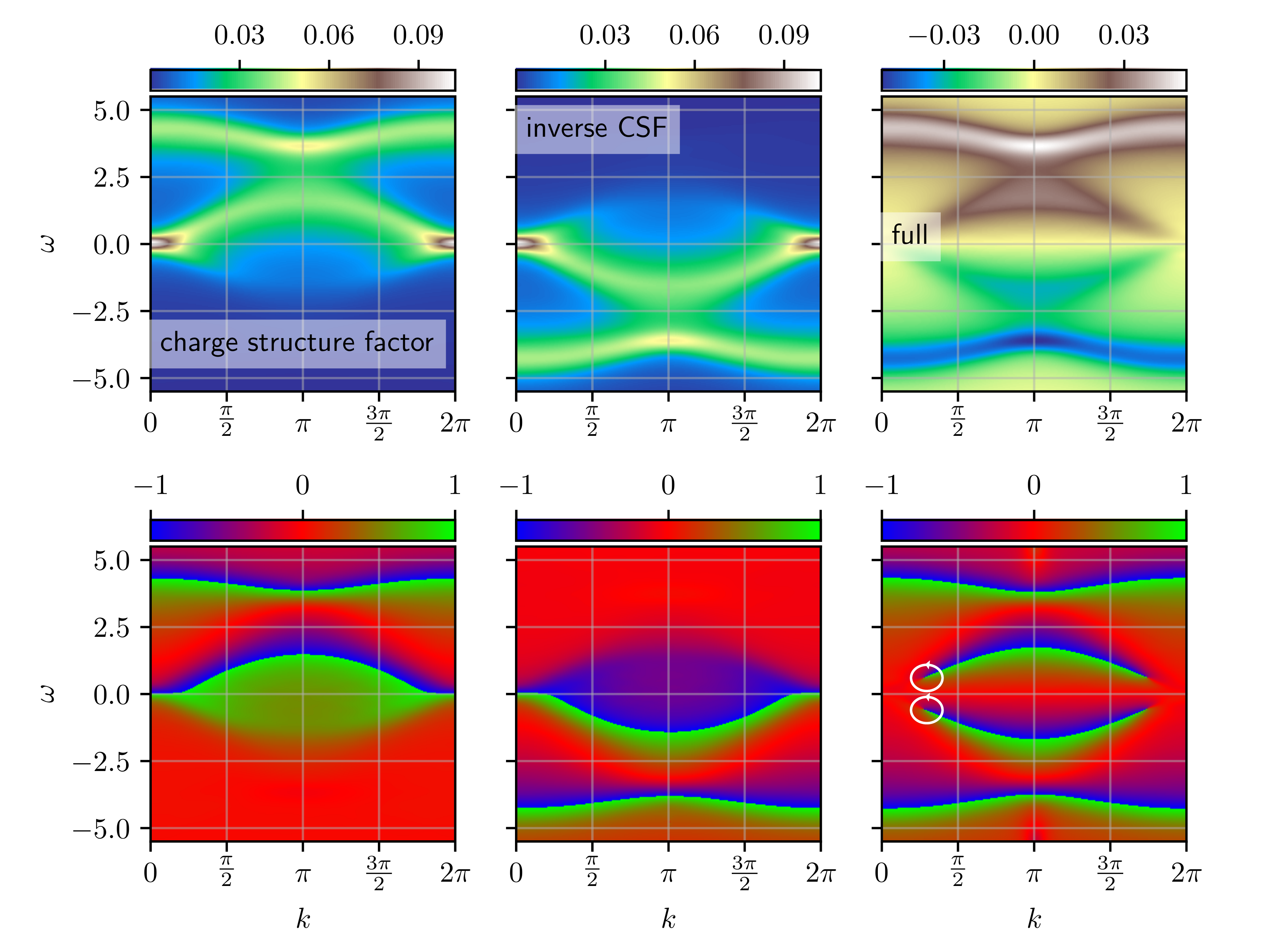}
\caption{\label{fig:2pTarg}
Top: Trace of the spectral function $-1/\pi \mathrm{Im} \mathrm{Tr} G^{cc}\left(\omega,k\right)$ from the two-particle Green's function Eq.~(\ref{eq:Gcc2}) after the Fourier transform Eq.~(\ref{eq:G_FT}). The left panel shows the charge structure factor only (CSF, first term), the center panel shows the inverse charge structure factor only (ICSF, second term). The right panel shows the full spectral function (commutator of the two terms).
Bottom: The corresponding phase function $\phi\left(\omega,k\right)$, Eq.~(\ref{eq:arg}). The white circles illustrate finite vorticity (\ref{eq:vort}) around two of the four exceptional points.
Parameters: $U_A=4$, $U_B=0$, temperature $T=1$. See appendix~\ref{app:Methods} for further technical details.
}
\end{center}
\end{figure*}

\begin{figure*}[!htb]
\centering
\begin{center}
\includegraphics[width=0.5\textwidth]{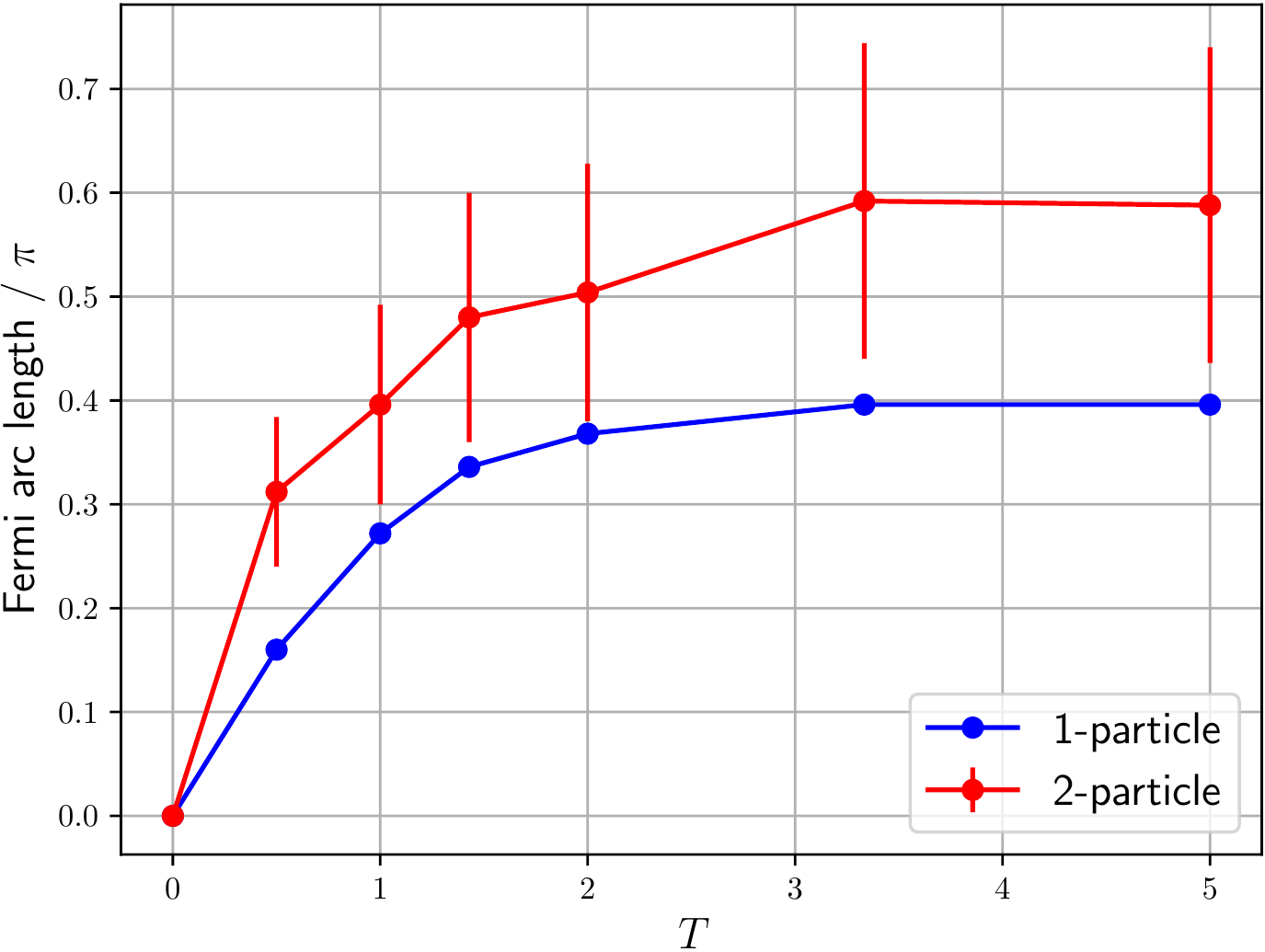}
\caption{\label{fig:arcLength}
Length of the one- and two-particle Fermi arc as a function of temperature for $U_A=4$, $U_B=0$. The one-particle Fermi arc is defined to lie between the exceptional points (cf. figure~\ref{fig:1pT}). The two-particle Fermi arc is read off from the spectral function (cf. figure~\ref{fig:2pTarg}), defined by the points where the spectral weight drops by 50\% compared to the maximal value at $k=0$. The vertical bars correspond to the range 45\%-55\%.
}
\end{center}
\end{figure*}

Figure~\ref{fig:2pTarg} shows the results, which one can compare with $T=0$ in the right panels of figure~\ref{fig:2_T0}. The CSF part at $T=0$ has a two-band structure: a gapless band that touches $\omega=0$ around $k=0$; and a gapped band with little dispersion whose gap grows with $U_A$. We surmise that the latter is interpretable as a band of paired electrons (``doublons''~\cite{Rausch_2016}). At finite temperature, we observe a strong shift of the spectral weight from the gapped band to $\omega=0$, $k=0$ in what can be indeed called a ``two-particle Fermi arc''. Figure~\ref{fig:arcLength} indicates that it is about 1.5 as large as the corresponding one-particle arc at a given temperature $T$.

In the noninteracting case, the two-particle Green's function is given by the Lindhard formula, which for a multiband system reads
\begin{equation}
\begin{split}
G^{cc}_{\mu\nu}\left(\omega,k\right) &= \frac{1}{2}\sum_{q}\sum_{ss'} M^{ss'}_{\mu\nu}\left(q,q+k\right) \\
&\quad\times\frac{\left<n_{qs}\right>-\left<n_{k+q,s'}\right>}{\omega+i0^+-\epsilon_{s'}\left(q+k\right)+\epsilon_s\left(q\right)},
\label{eq:Lindhard}
\end{split}
\end{equation}
where the matrix element $M^{ss'}_{\mu\nu}\left(q,q+k\right)$ is related to the eigenvectors $v_{s\mu}\left(k\right)$ of the unit cell Hamiltonian via
\begin{equation}
M^{ss'}_{\mu\nu}\left(q,q+k\right) = v_{s\mu}\left(q\right) v^*_{s'\mu}\left(k+q\right) v_{s'\nu}\left(k+q\right) v^*_{s\nu}\left(q\right).
\end{equation}
In our case, we can label the two bands by a sign $s=\pm$ and explicitly have $v_{\pm}\left(k\right)=1/\sqrt{2}\left[1,\pm\exp\left(-ik/2\right)\right]$ and $\epsilon_{\pm}\left(k\right) = \pm 2\cos\left(k/2\right)$. The temperature-dependent occupation numbers $\left<n_{ks}\right>$ are given by the Fermi function.
Due to the presence of the additional momentum summation, the formula is difficult to analyze even in the noninteracting case. However, we note that it has a self-convolution form, whereby single-particle properties like the bandwidth are expected to double in size~\cite{Rausch_2016}, thus offering an intuitive explanation for the larger Fermi arc. It may also explain the appearance of the Fermi arc at $k=0$, since the one-particle momenta are added up via $k=\left(\pi+\pi\right)\textrm{mod}~2\pi=0$. 
Even though the interacting $G^{cc}$ is of course \textit{not} given by a mere self-convolution, we may be seeing a very similar behavior due to the presence of noninteracting sites. Note that the one-particle spectral function also shows such a mixed behavior, with renormalized free-particle bands \textit{in addition} to upper and lower Hubbard bands, but not replaced by them as in the homogeneous case.

Looking at the full $G^{cc}$, we find that the two-particle Fermi arc vanishes, as it is canceled due to the relative minus sign (right part of figure~\ref{fig:2pTarg}). This is a general property of a commutator Green's function, which fulfills $G^*_{XY}\left(t\right)=G_{X^{\dagger}Y^{\dagger}}\left(t\right)=G_{XY}\left(t\right)$ if $X=X^{\dagger}$ and $Y=Y^{\dagger}$, so that the imaginary part at $\omega=0$ (i.e. the integral over $t$) has to vanish.

An important thing to note is that the chiral many-body symmetry of the Hamiltonian results in a different symmetry constraint for the two-particle Green's function as compared to the single-particle one. We find that it leads to the following relation for $G^{cc}$ (the derivation is outlined in appendix~\ref{app:Gcc_symm}):
\begin{equation}
G^{cc}_{\mu\nu}\left(\omega,k\right) = \left(\left[G^{cc}\left(-\omega,k\right)\right]^{\dagger}\right)_{\mu\nu}.
\label{eq:Gcc_symm}
\end{equation}
Thus, we now should look at the vorticity, Eq.~(\ref{eq:arg}), with $G^{1p}$ replaced by $G^{cc}$. Interestingly, we do not find any points where either the CSF or ICSF part becomes defective (non-diagonalizable), but such points appear in the full $G^{cc}$. This is displayed in the lower panels of figure~\ref{fig:2pTarg}. The phase discontinuities of $G^{cc}$ now have endpoints, where a finite vorticity is picked up along closed integration loops (indicated by white circles). By looking at the eigenvalues of the matrix $G^{cc}$, we can confirm that it becomes defective at the endpoints, as far as finite numerics allows us to say it.

These endpoints result from the fact that the phase discontinuities exactly touch $\omega=0$ at the Fermi arc in both the CSF and the ICSF part, leading to an exact cancellation in the commutator.  In this way, two pairs of exceptional points appear, and they move further away from $k=0$ with increasing temperature (we show just one temperature point for brevity). An intriguing effect is that they are also not confined to $\omega=0$ due to the distinct symmetry constraint for the two-particle Green's function. However, the precise relation between the exceptional points of $G^{cc}$ and the Fermi arc in the CSF part is not clear and remains an interesting open question that is left for future investigations.

%%%%%%%%%%%%%%%%%%%%%%%%%%%%%
\section{\label{sec:Summary}Discussion}
%%%%%%%%%%%%%%%%%%%%%%%%%%%%%

We have demonstrated the existence of novel non-Hermitian effects that were recently predicted to appear in a one-dimensional chiral-symmetric system based on symmetry considerations~\cite{Yoshida_2020}: A Dirac point of the noninteracting band structure at the Fermi energy splits into two exceptional points, with a 1D Fermi arc (flat band) in between, when sublattice-dependent interactions at finite temperature are introduced. This is both a dramatic effect of electron-electron correlations and of finite temperature in 1D that goes beyond the mere smearing out of the spectral features. The exceptional points are to a large degree robust against perturbations, such as hopping dimerization, which do not break the chiral symmetry. They are probably even robust against small symmetry-breaking perturbations, but have to move to finite values of $\omega$.

Examining the two-particle charge-charge spectral function, we find a Fermi arc that is roughly 1.5 as large as in the one-particle case when restricting ourselves to just the charge structure factor or its time-inverse counterpart. However, when looking at the full-commutator Greens' function, we find that two pairs of exceptional points appear, while the Fermi arc at $\omega=0$ is canceled out. Furthermore, they appear away from $\omega=0$ as endpoints in the discontinuity line of the phase function Eq.~(\ref{eq:arg}). The large spectral weight at $\omega=0$ in this case is not related to long-range order, which is suppressed by the finite temperature.

The two-particle spectral function is crucial in characterizing an interacting many-body system, but is much more challenging to analyze, as its noninteracting form is already not simple and an effective Hamiltonian cannot be easily defined. Our data showing that exceptional points still exist suggests that interesting non-Hermitian effects may still be waiting to be discovered, and can hopefully stimulate further studies.

In particular, recent advances in the calculation of one- and two-particle spectral functions for strongly correlated 2D systems~\cite{Tanaka_2019,Krien_2020} may allow to extend the study to the highly interesting field of 2D physics, where chiral symmetry leads to exceptional rings in the one-particle spectrum~\cite{Yoshida_2019,Yoshida_2020}.

Since the novel non-Hermitian effects persist in various spectral functions, experimentally this offers a wide array of possibilities to access them. A prime candidate would be angle-resolved photoemission (ARPES) for the one-particle case, while Bragg spectroscopy should in principle be able to measure the charge structure factor close to $\omega=0$. Because of the charge-SU(2) symmetry (see appendix~\ref{app:Unitary}), Auger spectroscopy may also be considered. In all cases, one should look for the flat band of the 1D Fermi arc.
Superlattices of correlated and non-correlated materials may provide material candidates~\cite{Baibich_1988}. Another possibility would be 1D optical lattices with controlled spatially modulated interactions~\cite{Clark_2015, Arunkumar_2019}.

%%%%%%%%%%%%%%%
\begin{acknowledgments}
%%%%%%%%%%%%%%%
R.R. thanks the Japan Society for the Promotion of Science (JSPS) and the Alexander von Humboldt Foundation. Computations were partially performed at the Yukawa Institute for Theoretical Physics, Kyoto. R.R. gratefully acknowledges support by JSPS, KAKENHI Grant No. JP18F18750. R.P. is supported by  JSPS, KAKENHI Grant No. JP18K03511. T.Y. is supported by JSPS, KAKENHI Grants No.~JP19K21032 and No.~JP20H04627.
\end{acknowledgments}

\clearpage

\appendix

%%%%%
\section{\label{app:Methods}Methods}
%%%%%

For the $T=0$ spectral functions in figure \ref{fig:2_T0}, we determine the ground state of the infinite chain by using the \textit{variational uniform matrix-product state} (VUMPS) framework~\cite{Zauner-Stauber_2018}. The Green's function is calculated for infinite boundary conditions~\cite{Phien_2012} using a real-time propagation algorithm based on the time-dependent variational principle (TDVP)~\cite{Haegeman_2016}. The local perturbation is allowed to spread on an inhomogeneous segment of $L=104$ sites, with a time step of $dt=0.1$, up to a maximal propagation time of $t\textsubscript{max}=48$ inverse hoppings. This cutoff time merely affects the resolution of the spectrum and does not neglect any spectral features.

Finite temperatures are incorporated into the matrix-product state framework using standard techniques~\cite{Feiguin_2005, Karrasch_2013, Barthel_2016, Nocera_2016}: By doubling the system's degrees of freedom we effectively go to a description using a density matrix which is purified into a vector. We initiate the $\beta=1/T=0$ state on a finite chain of $L=64$ sites and propagate up to the desired value of $\beta$ with a step size of $d\beta=0.1$ using the two-site TDVP algorithm~\cite{Haegeman_2016}. Applying the local perturbation to the resulting finite-temperature state, we are then able to propagate in real time up to a maximal cutoff value of $t\textsubscript{max}=16$ (time steps: $dt=0.025$ for the single-particle case, $dt=0.1$ for the two-particle case). Since finite temperature introduces a natural broadening, the spectra converge with respect to $t\textsubscript{max}$ and further propagation is not necessary. The growth of the entanglement entropy can be kept in check by counterpropagating the bath sites~\cite{Karrasch_2013}. 

%%%%%%%%%%%%%%%%%%%%%%%%
\section{\label{app:Sigma}Momentum dependence of self-energy}
%%%%%%%%%%%%%%%%%%%%%%%%

\begin{figure*}[!htb]
\centering
\begin{center}
\includegraphics[width=0.8\textwidth]{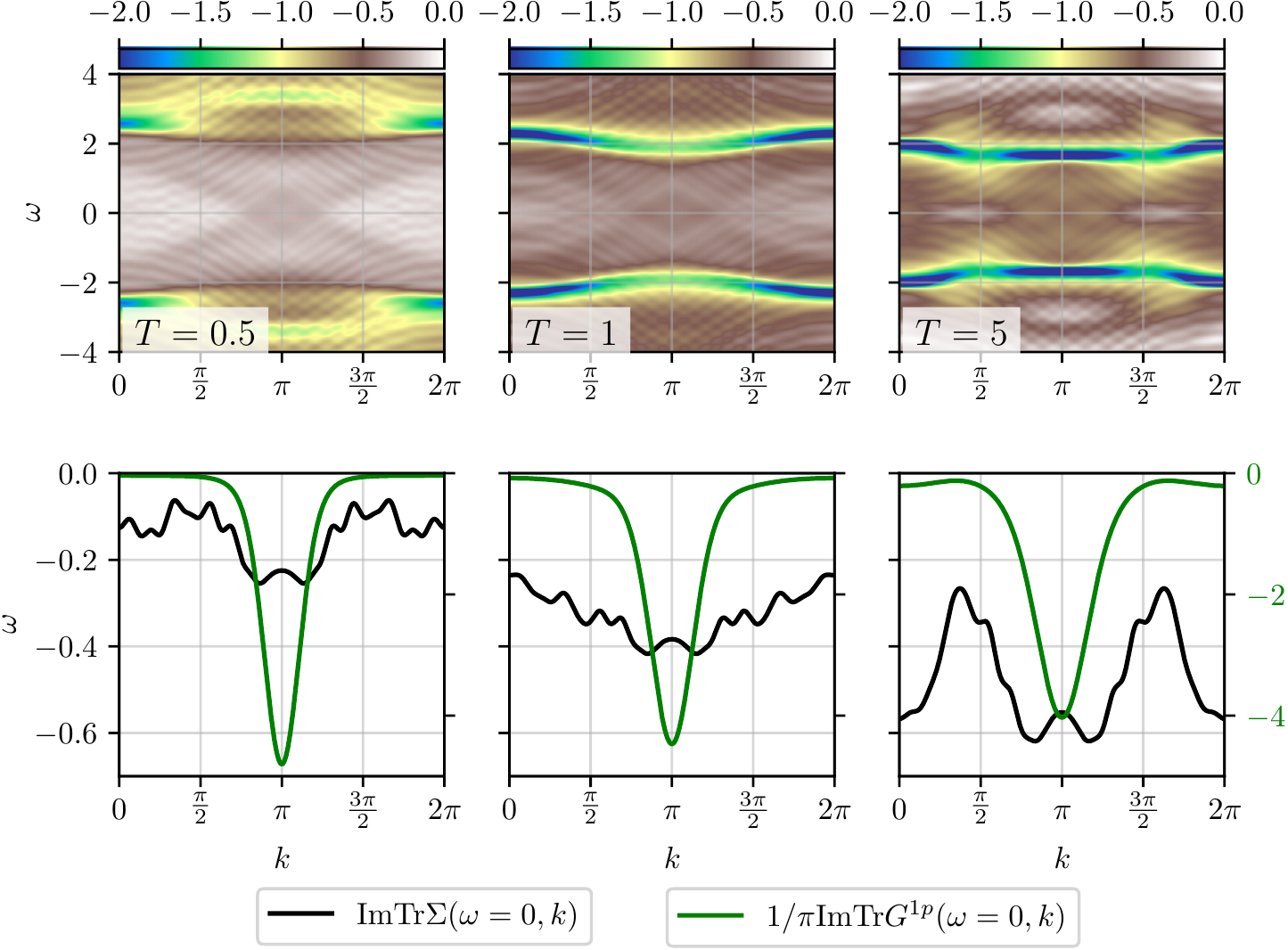}
\caption{\label{fig:SigmaT}
Top: Trace of the momentum-resolved imaginary part of the self-energy $\mathrm{Im} \mathrm{Tr} \Sigma\left(\omega,k\right)$ calculated at the temperatures $T=0.5,1,5$ (see appendix~\ref{app:Methods}). Bottom: Trace of the momentum-resolved self-energy and of the negative one-particle spectral function $1/\pi \mathrm{Im} \mathrm{Tr} G^{1p}\left(\omega=0,k\right)$ at the Fermi energy for the same temperatures.
}
\end{center}
\end{figure*}

By inverting the $2\times2$ matrix of one-particle Green's function with and without interaction for each value of $\omega$ and $k$, we are able to calculate the momentum-resolved self-energy $\Sigma\left(\omega,k\right)=\left[G^{1p}_0\left(\omega,k\right)\right]^{-1}-\left[G^{1p}\left(\omega,k\right)\right]^{-1}$, shown in figure~\ref{fig:SigmaT}.

We notice that the imaginary part of the self-energy is peaked around the Hubbard bands $\omega\approx\pm U/2 \sim \pm U$ and decreases towards $\omega=0$. Its overall value at $\omega=0$ grows with temperature. 
We also note that the self-energy strongly depends on the momentum $k$ (particularly at low temperatures), a signature of the strong spatial fluctuations in 1D. Any approximation that neglects the momentum dependence of $\Sigma\left(\omega,k\right)$ would thus be insufficient.

%%%%%%%%%%%%%%%%%%%%%%%%
\section{\label{app:Gcc_symm}Symmetry constraint of the 2-particle Green's function}
%%%%%%%%%%%%%%%%%%%%%%%%

The many-body chiral symmetry $U$ acts on the creation and annihilation operators in the following way~\cite{Gurarie_2011,Yoshida_2019}:
\begin{eqnarray}
U^{\dagger}c_{m\mu\sigma}U &= \sum_{\nu} c^{\dagger}_{m\nu\sigma} \left(u^{\dagger}\right)_{\nu\mu},\\
U^{\dagger}c^{\dagger}_{m\mu\sigma}U &= \sum_{\nu} u_{\mu\nu} c_{m\nu\sigma},
\end{eqnarray}
with $u_{\mu\nu}=\left(\tau^z\right)_{\mu\nu}$. This implies that the shifted density transforms as:
\begin{equation}
U^{\dagger}\left(n_{m\mu\sigma}-1/2\right)U= -\left(n_{m\mu\sigma}-1/2\right).
\end{equation}
Proceeding in the same way as in the above references and plugging the transformation into the density-density correlator, we find
\begin{equation}
\left<Q^z_{m\mu}\left(t\right)Q^z_{n\nu}\right>\theta\left(t\right) = \left<Q^z_{n\nu}Q^z_{m\mu}\left(-t\right)\right>\theta\left(t\right),
\end{equation}
with the z-component of the pseudospin from (\ref{suppleq:Qz}). Defining the retarded and advanced commutator Green's functions for two operators $X$ and $Y$ in the standard way,
\begin{eqnarray}
G^{\textsuperscript{ret}}_{XY}\left(t\right) &= -i\theta\left(+t\right) \left<\big[X\left(t\right),Y\big]\right>, \\
G^{\textsuperscript{adv}}_{XY}\left(t\right) &= +i\theta\left(-t\right) \left<\big[X\left(t\right),Y\big]\right>,
\end{eqnarray}
and using
\begin{equation}
c_{\mu\sigma}\left(k\right) = \frac{1}{\sqrt{L/L_c}} \sum_m e^{-ikmL_c} c_{m\mu\sigma},
\end{equation}
so that
\begin{equation}
Q^z_{\mu}\left(k\right) = 1/2 \left(\sum_{k'\sigma} c^{\dagger}_{k'\mu\sigma}c_{k+k',\mu\sigma}-1\right),
\end{equation}
we find the following relation:
\begin{equation}
G^{cc}\textsuperscript{,ret}_{\mu\nu}\left(\omega,k\right) = G^{cc}\textsuperscript{,adv}_{\mu\nu}\left(-\omega,k\right).
\label{eq:Gra1}
\end{equation}
To eliminate the appearance of the advanced Green's function, we use the following formula that follows from the definition of the Green's function and Hermiticity of the density operators:
\begin{equation}
\left[G^{cc}\textsuperscript{,ret}_{\mu\nu}\left(\omega,k\right)\right]^* = G^{cc}\textsuperscript{,adv}_{\nu\mu}\left(\omega,k\right).
\label{eq:Gra2}
\end{equation}
Combining (\ref{eq:Gra1}) and (\ref{eq:Gra2}), we obtain the relation presented in the main text:
\begin{equation}
G^{cc}\textsuperscript{,ret}_{\mu\nu}\left(\omega,k\right) = \left(\left[G^{cc}\textsuperscript{,ret}\left(-\omega,k\right)\right]^{\dagger}\right)_{\mu\nu}.
\end{equation}

%%%%%%%%%%%%%%%%%%%%%%%%
\section{\label{app:Unitary}Unitary Symmetries}
%%%%%%%%%%%%%%%%%%%%%%%%

Our model possesses both the spin-SU(2) and charge-SU(2) symmetry~\cite{Yang_1990, Essler_2005, Anderson_1958}.

The spin-SU(2) is given by $\left[H,\sum_i\ensuremath{\mathbf{S}}_i\right]=0$ with the spin vector $\ensuremath{\mathbf{S}}_i=\left(S^x_i,S^y_i,S^z_i\right)$, whose z-component is given by 
\begin{equation}
S^z_i=1/2\left(n_{i\uparrow}-n_{i\downarrow}\right),
\end{equation}
and the x- and y-components are given by $S^x_i=1/2\left(S^+_i+S^-_i\right)$, $S^y_i=1/2i\left(S^+_i-S^-_i\right)$, with the ladder operators
\begin{equation}
S^+_i = c^{\dagger}_{i\uparrow} c_{i\downarrow}
\end{equation}
and $S^-_i=\left(S^+_i\right)^{\dagger}$.

The charge-SU(2) is given by $\left[H,\sum_i\ensuremath{\mathbf{Q}}_i\right]=0$ with the pseudospin vector  $\ensuremath{\mathbf{Q}}_i=\left(Q^x_i,Q^y_i,Q^z_i\right)$, whose z-component is given by 
\begin{equation}
Q^z_i=1/2\left(n_{i}-1\right),
\label{suppleq:Qz}
\end{equation}
and the x- and y-components are given by $Q^x_i=1/2\left(Q^+_i+Q^-_i\right)$, $Q^y_i=1/2i\left(Q^+_i-Q^-_i\right)$, with the ladder operators \begin{equation}
Q^+_i=\left(-1\right)^ic_{i\uparrow}c_{i\downarrow}
\end{equation}
and $Q^-_i=\left(Q^+_i\right)^{\dagger}$.

In the general case, we can choose out of six local two-particle excitation operators, namely $S^+_i$, $S^-_i$, $S^z_i$, $Q^+_i$, $Q^-_i$, $Q^z_i$. However, due to the SU(2) symmetries, all components of $\ensuremath{\mathbf{S}}_i$ on the one hand, and of $\ensuremath{\mathbf{Q}}_i$ on the other hand are equivalent, so that we can restrict ourselves to an analysis of $S^z_i$ (spin-spin) and $Q^z_i$ (charge-charge). Thus, the dynamic charge structure factor (excitations by $Q_i^z$) is related to the Auger spectral function (excitations by $\left(-1\right)^iQ_{i}^+$)~\cite{Rausch_2016}, and either one can be measured experimentally to observe the effects described in the main text.

\section*{References}
\bibliographystyle{iopart-num}
%\bibliography{nHerm1D}

\begin{thebibliography}{10}
\expandafter\ifx\csname url\endcsname\relax
  \def\url#1{{\tt #1}}\fi
\expandafter\ifx\csname urlprefix\endcsname\relax\def\urlprefix{URL }\fi
\providecommand{\eprint}[2][]{\url{#2}}

\bibitem{Fukui_1998}
Fukui T and Kawakami N 1998 {\em Physical Review B\/} {\bf 58} 16051

\bibitem{Ashida_2020}
Ashida Y, Gong Z and Ueda M 2020 Non-hermitian physics (\textit{Preprint}
  \eprint{2006.01837})

\bibitem{Feng_2017}
Feng L, El-Ganainy R and Ge L 2017 {\em Nature Photonics\/} {\bf 11} 752

\bibitem{Ozawa_2019}
Ozawa T, Price H~M, Amo A, Goldman N, Hafezi M, Lu L, Rechtsman M~C, Schuster
  D, Simon J, Zilberberg O and Carusotto I 2019 {\em Rev. Mod. Phys.\/} {\bf
  91}(1) 015006
  \urlprefix\url{https://link.aps.org/doi/10.1103/RevModPhys.91.015006}

\bibitem{Oezdemir_2019}
{\"O}zdemir {\c S}~K, Rotter S, Nori F and Yang L 2019 {\em Nature Materials\/}
  {\bf 18} 783--798 \urlprefix\url{https://doi.org/10.1038/s41563-019-0304-9}

\bibitem{Yamamoto_2019}
Yamamoto K, Nakagawa M, Adachi K, Takasan K, Ueda M and Kawakami N 2019 {\em
  Phys. Rev. Lett.\/} {\bf 123}(12) 123601
  \urlprefix\url{https://link.aps.org/doi/10.1103/PhysRevLett.123.123601}

\bibitem{Yoshida_2020c}
Yoshida T, Kudo K, Katsura H and Hatsugai Y 2020 {\em arXiv preprint
  arXiv:2005.12635\/}

\bibitem{Kozii_2017}
Kozii V and Fu L 2017 Non-hermitian topological theory of finite-lifetime
  quasiparticles: Prediction of bulk fermi arc due to exceptional point
  (\textit{Preprint} \eprint{1708.05841})

\bibitem{Yoshida_2018}
Yoshida T, Peters R and Kawakami N 2018 {\em Phys. Rev. B\/} {\bf 98}(3) 035141
  \urlprefix\url{https://link.aps.org/doi/10.1103/PhysRevB.98.035141}

\bibitem{Kimura_2019}
Kimura K, Yoshida T and Kawakami N 2019 {\em Phys. Rev. B\/} {\bf 100}(11)
  115124 \urlprefix\url{https://link.aps.org/doi/10.1103/PhysRevB.100.115124}

\bibitem{Matsushita_2019}
Matsushita T, Nagai Y and Fujimoto S 2019 {\em Phys. Rev. B\/} {\bf 100}(24)
  245205 \urlprefix\url{https://link.aps.org/doi/10.1103/PhysRevB.100.245205}

\bibitem{Michishita_2020}
Michishita Y and Peters R 2020 {\em Phys. Rev. Lett.\/} {\bf 124}(19) 196401
  \urlprefix\url{https://link.aps.org/doi/10.1103/PhysRevLett.124.196401}

\bibitem{Michishita_2020b}
Michishita Y, Yoshida T and Peters R 2020 {\em Phys. Rev. B\/} {\bf 101}(8)
  085122 \urlprefix\url{https://link.aps.org/doi/10.1103/PhysRevB.101.085122}

\bibitem{Yoshida_2020}
Yoshida T, Peters R, Kawakami N and Hatsugai Y 2020 Exceptional band touching
  for strongly correlated systems in equilibrium (\textit{Preprint}
  \eprint{2002.11265})

\bibitem{Zyuzin_2018}
Zyuzin A~A and Zyuzin A~Y 2018 {\em Phys. Rev. B\/} {\bf 97}(4) 041203
  \urlprefix\url{https://link.aps.org/doi/10.1103/PhysRevB.97.041203}

\bibitem{Shen_2018}
Shen H and Fu L 2018 {\em arXiv preprint arXiv:1802.03023\/}

\bibitem{Papaj_2019}
Papaj M, Isobe H and Fu L 2019 {\em Phys. Rev. B\/} {\bf 99}(20) 201107
  \urlprefix\url{https://link.aps.org/doi/10.1103/PhysRevB.99.201107}

\bibitem{Matsushita_2020}
Matsushita T, Nagai Y and Fujimoto S 2020 {\em arXiv preprint
  arXiv:2004.11014\/}

\bibitem{Shen_2018b}
Shen H, Zhen B and Fu L 2018 {\em Phys. Rev. Lett.\/} {\bf 120}(14) 146402
  \urlprefix\url{https://link.aps.org/doi/10.1103/PhysRevLett.120.146402}

\bibitem{Okugawa_2019}
Okugawa R and Yokoyama T 2019 {\em Phys. Rev. B\/} {\bf 99}(4) 041202
  \urlprefix\url{https://link.aps.org/doi/10.1103/PhysRevB.99.041202}

\bibitem{Budich_2019}
Budich J~C, Carlstr\"om J, Kunst F~K and Bergholtz E~J 2019 {\em Phys. Rev.
  B\/} {\bf 99}(4) 041406
  \urlprefix\url{https://link.aps.org/doi/10.1103/PhysRevB.99.041406}

\bibitem{Kawabata_2019b}
Kawabata K, Shiozaki K, Ueda M and Sato M 2019 {\em Phys. Rev. X\/} {\bf 9}(4)
  041015 \urlprefix\url{https://link.aps.org/doi/10.1103/PhysRevX.9.041015}

\bibitem{SanJose_2016}
San-Jose P, Cayao J, Prada E and Aguado R 2016 {\em Scientific reports\/} {\bf
  6} 21427

\bibitem{Gong_2018}
Gong Z, Ashida Y, Kawabata K, Takasan K, Higashikawa S and Ueda M 2018 {\em
  Phys. Rev. X\/} {\bf 8}(3) 031079
  \urlprefix\url{https://link.aps.org/doi/10.1103/PhysRevX.8.031079}

\bibitem{Bergholtz_2019}
Bergholtz E~J, Budich J~C and Kunst F~K 2019 {\em arXiv preprint
  arXiv:1912.10048\/}

\bibitem{Yoshida_2019c}
Yoshida T, Kudo K and Hatsugai Y 2019 {\em Scientific Reports\/} {\bf 9} 16895
  ISSN 2045-2322 \urlprefix\url{https://doi.org/10.1038/s41598-019-53253-8}

\bibitem{Liu_2020}
Liu T, He J~J, Yoshida T, Xiang Z~L and Nori F 2020 {\em arXiv preprint
  arXiv:2001.09475\/}

\bibitem{Yao_2018}
Yao S and Wang Z 2018 {\em Phys. Rev. Lett.\/} {\bf 121}(8) 086803
  \urlprefix\url{https://link.aps.org/doi/10.1103/PhysRevLett.121.086803}

\bibitem{Kunst_2018}
Kunst F~K, Edvardsson E, Budich J~C and Bergholtz E~J 2018 {\em Phys. Rev.
  Lett.\/} {\bf 121}(2) 026808
  \urlprefix\url{https://link.aps.org/doi/10.1103/PhysRevLett.121.026808}

\bibitem{Zhang_2019}
Zhang K, Yang Z and Fang C 2019 {\em arXiv preprint arXiv:1910.01131\/}

\bibitem{Lee_2019}
Lee C~H and Thomale R 2019 {\em Phys. Rev. B\/} {\bf 99}(20) 201103
  \urlprefix\url{https://link.aps.org/doi/10.1103/PhysRevB.99.201103}

\bibitem{Okuma_2020}
Okuma N, Kawabata K, Shiozaki K and Sato M 2020 {\em Phys. Rev. Lett.\/} {\bf
  124}(8) 086801
  \urlprefix\url{https://link.aps.org/doi/10.1103/PhysRevLett.124.086801}

\bibitem{Yokomizo_2019}
Yokomizo K and Murakami S 2019 {\em Phys. Rev. Lett.\/} {\bf 123}(6) 066404
  \urlprefix\url{https://link.aps.org/doi/10.1103/PhysRevLett.123.066404}

\bibitem{Borgnia_2020}
Borgnia D~S, Kruchkov A~J and Slager R~J 2020 {\em Phys. Rev. Lett.\/} {\bf
  124}(5) 056802
  \urlprefix\url{https://link.aps.org/doi/10.1103/PhysRevLett.124.056802}

\bibitem{Helbig_2020}
Helbig T, Hofmann T, Imhof S, Abdelghany M, Kiessling T, Molenkamp L~W, Lee
  C~H, Szameit A, Greiter M and Thomale R 2020 {\em Nature Physics\/} ISSN
  1745-2481 \urlprefix\url{https://doi.org/10.1038/s41567-020-0922-9}

\bibitem{Hofmann_2020}
Hofmann T, Helbig T, Schindler F, Salgo N, Brzezi\ifmmode~\acute{n}\else
  \'{n}\fi{}ska M, Greiter M, Kiessling T, Wolf D, Vollhardt A,
  Kaba\ifmmode~\check{s}\else \v{s}\fi{}i A, Lee C~H, Bilu\ifmmode
  \check{s}\else \v{s}\fi{}i\ifmmode~\acute{c}\else \'{c}\fi{} A, Thomale R and
  Neupert T 2020 {\em Phys. Rev. Research\/} {\bf 2}(2) 023265
  \urlprefix\url{https://link.aps.org/doi/10.1103/PhysRevResearch.2.023265}

\bibitem{Xiao_2020}
Xiao L, Deng T, Wang K, Zhu G, Wang Z, Yi W and Xue P 2020 {\em Nature
  Physics\/} ISSN 1745-2481
  \urlprefix\url{https://doi.org/10.1038/s41567-020-0836-6}

\bibitem{Yoshida_2020b}
Yoshida T, Mizoguchi T and Hatsugai Y 2020 {\em Phys. Rev. Research\/} {\bf
  2}(2) 022062
  \urlprefix\url{https://link.aps.org/doi/10.1103/PhysRevResearch.2.022062}

\bibitem{Yoshida_2019}
Yoshida T, Peters R, Kawakami N and Hatsugai Y 2019 {\em Phys. Rev. B\/} {\bf
  99}(12) 121101
  \urlprefix\url{https://link.aps.org/doi/10.1103/PhysRevB.99.121101}

\bibitem{Paiva_1996}
Paiva T and dos Santos R~R 1996 {\em Phys. Rev. Lett.\/} {\bf 76}(7) 1126--1129
  \urlprefix\url{https://link.aps.org/doi/10.1103/PhysRevLett.76.1126}

\bibitem{Li_2018}
Li J, Cheng C, Paiva T, Lin H~Q and Mondaini R 2018 {\em Phys. Rev. Lett.\/}
  {\bf 121}(2) 020403
  \urlprefix\url{https://link.aps.org/doi/10.1103/PhysRevLett.121.020403}

\bibitem{Baibich_1988}
Baibich M~N, Broto J~M, Fert A, Van~Dau F~N, Petroff F, Etienne P, Creuzet G,
  Friederich A and Chazelas J 1988 {\em Phys. Rev. Lett.\/} {\bf 61}(21)
  2472--2475
  \urlprefix\url{https://link.aps.org/doi/10.1103/PhysRevLett.61.2472}

\bibitem{Paiva_1998}
Paiva T and dos Santos R~R 1998 {\em Phys. Rev. B\/} {\bf 58}(15) 9607--9610
  \urlprefix\url{https://link.aps.org/doi/10.1103/PhysRevB.58.9607}

\bibitem{Paiva_2002}
Paiva T and dos Santos R~R 2002 {\em Phys. Rev. B\/} {\bf 65}(15) 153101
  \urlprefix\url{https://link.aps.org/doi/10.1103/PhysRevB.65.153101}

\bibitem{Malvezzi_2006}
Malvezzi A~L, Paiva T and dos Santos R~R 2006 {\em Phys. Rev. B\/} {\bf 73}(19)
  193407 \urlprefix\url{https://link.aps.org/doi/10.1103/PhysRevB.73.193407}

\bibitem{Zhang_2015}
Zhang L~L, Huang J, Duan C~B and Wang W~Z 2015 {\em Modern Physics Letters B\/}
  {\bf 29} 1550208 (\textit{Preprint}
  \eprint{https://doi.org/10.1142/S0217984915502085})
  \urlprefix\url{https://doi.org/10.1142/S0217984915502085}

\bibitem{Feiguin_2005}
Feiguin A~E and White S~R 2005 {\em Phys. Rev. B\/} {\bf 72}(22) 220401
  \urlprefix\url{https://link.aps.org/doi/10.1103/PhysRevB.72.220401}

\bibitem{Karrasch_2013}
Karrasch C, Bardarson J~H and Moore J~E 2013 {\em New Journal of Physics\/}
  {\bf 15} 083031
  \urlprefix\url{https://doi.org/10.1088%2F1367-2630%2F15%2F8%2F083031}

\bibitem{Barthel_2016}
Barthel T 2016 {\em Phys. Rev. B\/} {\bf 94}(11) 115157
  \urlprefix\url{https://link.aps.org/doi/10.1103/PhysRevB.94.115157}

\bibitem{Nocera_2016}
Nocera A and Alvarez G 2016 {\em Phys. Rev. B\/} {\bf 93}(4) 045137
  \urlprefix\url{https://link.aps.org/doi/10.1103/PhysRevB.93.045137}

\bibitem{Kawabata_2019c}
Kawabata K, Bessho T and Sato M 2019 {\em Phys. Rev. Lett.\/} {\bf 123}(6)
  066405
  \urlprefix\url{https://link.aps.org/doi/10.1103/PhysRevLett.123.066405}

\bibitem{Rausch_2016}
Rausch R and Potthoff M 2016 {\em New Journal of Physics\/} {\bf 18} 023033
  \urlprefix\url{https://doi.org/10.1088%2F1367-2630%2F18%2F2%2F023033}

\bibitem{Clark_2015}
Clark L~W, Ha L~C, Xu C~Y and Chin C 2015 {\em Phys. Rev. Lett.\/} {\bf
  115}(15) 155301
  \urlprefix\url{https://link.aps.org/doi/10.1103/PhysRevLett.115.155301}

\bibitem{Arunkumar_2019}
Arunkumar N, Jagannathan A and Thomas J~E 2019 {\em Phys. Rev. Lett.\/} {\bf
  122}(4) 040405
  \urlprefix\url{https://link.aps.org/doi/10.1103/PhysRevLett.122.040405}

\bibitem{Tanaka_2019}
Tanaka A 2019 {\em Phys. Rev. B\/} {\bf 99}(20) 205133
  \urlprefix\url{https://link.aps.org/doi/10.1103/PhysRevB.99.205133}

\bibitem{Krien_2020}
Krien F, Valli A, Chalupa P, Capone M, Lichtenstein A~I and Toschi A 2020 {\em
  Phys. Rev. B\/} {\bf 102}(19) 195131
  \urlprefix\url{https://link.aps.org/doi/10.1103/PhysRevB.102.195131}

\bibitem{Zauner-Stauber_2018}
Zauner-Stauber V, Vanderstraeten L, Fishman M~T, Verstraete F and Haegeman J
  2018 {\em Phys. Rev. B\/} {\bf 97}(4) 045145
  \urlprefix\url{https://link.aps.org/doi/10.1103/PhysRevB.97.045145}

\bibitem{Phien_2012}
Phien H~N, Vidal G and McCulloch I~P 2012 {\em Phys. Rev. B\/} {\bf 86}(24)
  245107 \urlprefix\url{https://link.aps.org/doi/10.1103/PhysRevB.86.245107}

\bibitem{Haegeman_2016}
Haegeman J, Lubich C, Oseledets I, Vandereycken B and Verstraete F 2016 {\em
  Phys. Rev. B\/} {\bf 94}(16) 165116
  \urlprefix\url{https://link.aps.org/doi/10.1103/PhysRevB.94.165116}

\bibitem{Gurarie_2011}
Gurarie V 2011 {\em Phys. Rev. B\/} {\bf 83}(8) 085426
  \urlprefix\url{https://link.aps.org/doi/10.1103/PhysRevB.83.085426}

\bibitem{Yang_1990}
Yang C~N and Zhang S 1990 {\em Modern Physics Letters B\/} {\bf 04} 759--766
  (\textit{Preprint} \eprint{https://doi.org/10.1142/S0217984990000933})
  \urlprefix\url{https://doi.org/10.1142/S0217984990000933}

\bibitem{Essler_2005}
Essler F~H, Frahm H, G{\"o}hmann F, Kl{\"u}mper A and Korepin V~E 2005 {\em The
  one-dimensional Hubbard model\/} (Cambridge University Press)

\bibitem{Anderson_1958}
Anderson P~W 1958 {\em Phys. Rev.\/} {\bf 112}(6) 1900--1916
  \urlprefix\url{https://link.aps.org/doi/10.1103/PhysRev.112.1900}

\end{thebibliography}

\providecommand{\newblock}{}

\end{document}